\documentclass{article}
\usepackage[utf8]{inputenc} 
\usepackage[english]{babel} 
\usepackage{url}   
\usepackage[hidelinks]{hyperref}
\usepackage{booktabs}       
\usepackage{amsfonts}       
\usepackage{nicefrac}       
\usepackage{xcolor}
\usepackage{graphicx}
\usepackage{tabularx}
\usepackage{enumitem}
\usepackage{comment}
\usepackage{subcaption}
\usepackage{apacite}  
\begin{document}
\title{Mići Princ -- A Little Boy Teaching Speech Technologies the Chakavian Dialect}

\author{
  Nikola LJUBEŠIĆ$^{1,2,3}$, Peter RUPNIK$^{1}$, Tea PERINČIĆ$^{4}$
}
\date{September 2024}

\maketitle
\small{
  $^{1}$Jožef Stefan Institute, Ljubljana, Slovenia \par
  $^{2}$University of Ljubljana, Slovenia \par
  $^{3}$Institute for Contemporary History, Ljubljana, Slovenia \par
  $^{4}$Maritime and History Museum of Croatian Littoral, Rijeka, Croatia
}
\hypertarget{}{}
\begin{abstract}

This paper documents our efforts in releasing the printed and audio book of the translation of the famous novel \emph{The Little Prince} into the Chakavian dialect, as a computer-readable, AI-ready dataset, with the textual and the audio components of the two releases now aligned on the level of each written and spoken word. Our motivation for working on this release is multiple. The first one is our wish to preserve the highly valuable and specific content beyond the small editions of the printed and the audio book. With the dataset published in the CLARIN.SI repository, this content is from now on at the fingertips of any interested individual. The second motivation is to make the data available for various artificial-intelligence-related usage scenarios, such as the one we follow upon inside this paper already -- adapting the \texttt{Whisper-large-v3} open automatic speech recognition model, with decent performance on standard Croatian, to Chakavian dialectal speech. We can happily report that with adapting the model, the word error rate on the selected test data has being reduced to a half, while we managed to remove up to two thirds of the error on character level. We envision many more usages of this dataset beyond the set of experiments we have already performed, both on tasks of artificial intelligence research and application, as well as dialectal research. The third motivation for this release is our hope that this, now highly structured dataset, will be transformed into a digital online edition of this work, allowing individuals beyond the research and technology communities to enjoy the beauty of the message of the little boy in the desert, told through the spectacular prism of the Chakavian dialect.
\end{abstract}

\textbf{Keywords:} The Little Prince, Chakavian dialect, text and speech dataset, automatic speech recognition

\section{Introduction}

We have recently witnessed staggering improvements in processing language in both textual~\cite{zhao2023survey} and speech modality~\cite{radford2022whisper}. Regardless of these drastic improvements in performance, they are mostly directed at well-resourced languages in their standardised form, disregarding the dialectal variation~\cite{kantharuban-etal-2023-quantifying} present in both the textual, but especially the spoken modality of language.

Our language in focus in this paper, the Croatian language, a member of the western group of South Slavic languages, has recently obtained its first open, large, searchable spoken dataset, namely the ParlaSpeech-HR corpus~\cite{ljubesic-etal-2022-parlaspeech}, based on parliamentary proceedings recordings and transcripts, currently consisting of 3,061 hours of spoken material and linguistically processed transcripts~\cite{11356/1914}.\footnote{The corpus is searchable through the CLARIN.SI concordancers at\url{https://tinyurl.com/parlaspeech}.
} The two only earlier examples of open spoken datasets of Croatian language that have to be mentioned here, especially important for their pioneering efforts, are the Croatian Adult Spoken Language Corpus (HrAL)~\cite{kuvavc2016croatian}, 250,000 tokens in size, and the CCCL Croatian corpus of child language~\cite{kovavcevic2002croatian}, consisting of recordings and detailed transcriptions of speech of three children.

The only open dialectal dataset of Croatian we are aware of is the recent textual translation of the COPA commonsense reasoning benchmark into the Chakavian dialect of Žminj, part of the DIALECT-COPA benchmark set~\cite{ljubesic-etal-2024-dialect}. There have, however, not been any open spoken dialectal datasets of Croatian. Here we are changing this, by releasing a small and aesthetically pleasing dataset, the aligned audio and printed book of the translation of \emph{The Little Prince} into various Chakavian micro-dialects -- \emph{Mići Princ}. The contributions of this work are the following:

\begin{enumerate}
    \item We are constructing and releasing via a FAIR (findable, accessible, interoperable, and reusable) repository the first open dataset of dialectal speech of the Croatian language, with speech aligned to its transcripts.
    \item We are releasing the dataset both in a rich, verbatim format, but also adapted for automatic speeech recognition (ASR) experiments, with instances of up to 30 seconds long, ready to be used for adapting or evaluating various ASR systems.
    \item We are showcasing the usefulness even of such a small dataset for modern speech technologies by successfully adapting the Whisper ASR model to the Chakavian dialect.
    \item We are releasing the first ASR system capable of processing the Chakavian dialect, lowering the relative word error rate on unseen speakers for around 40\%.
    \item We are paving the road to a digital online release of the underlying work, which will make the beauty of the Chakavian dialect significantly more accessible to the wider audiences.
\end{enumerate}

This paper is structured as follows:
we present the Mići Princ book in Section~\ref{sec:mici_princ}, in Section~\ref{sec:processing_description} we describe how we compiled the word-aligned dataset, and discuss steps needed to transform it into an ASR-specific dataset. In Section~\ref{sec:final_encoding} the encoding of both datasets is explained in detail. In Section~\ref{sec:asr_experiments} our ASR model fine-tuning is presented and the results obtained are commented. We discuss some limitations of our approach in section~\ref{sec:limitations} and wrap up with conclusions (Section~\ref{sec:conclusions}).

\section{Origin of the Data -- The Mići Princ Book}
\label{sec:mici_princ}

\emph{Mići Princ}~\cite{mici_princ_the_book} is a translation of \emph{Le Petit Prince} into various Chakavian micro-dialects, released both as a printed book and an audio book. Its special distinction is that almost every character in the book uses a different micro-dialect, which was achieved by using numerous translators and voice actors. In total, 16~translators and 16~voice actors were involved in the process, representing the 16 different characters in the audio and the text book.

\begin{figure}
    \centering
    \includegraphics[width=0.9\textwidth,alt={A map of the northern Adriatic Sea in grayscale, with orange circles showing locations of various voice actors. Orange circle labelled Dilavac is situated on the northern side of the Cres island, Geograf is placed situated in the centre of the island of Krk, and a dual marker for Autor and Mići Princ is placed on the continent, between Rijeka and the border with Slovenia.}]{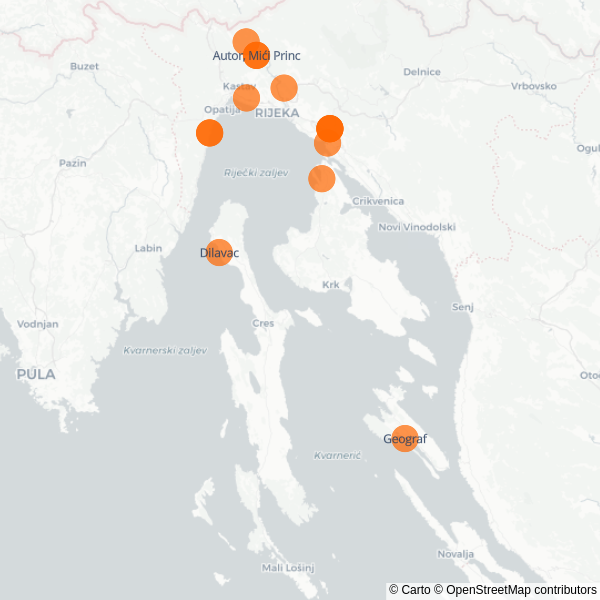}
    \caption{Origins of the voice actors, plotted on a map of northern Kvarner Gulf (Croatia). Speakers with labelled markers were used for model evaluation (see Section~\ref{sec:asr_experiments}).}
    \label{fig:map}
\end{figure}

The audio book spans 113~minutes, which also includes the music that is sometimes used to start or end a chapter. The duration of voiced segments only is 79~minutes. The text portion (after removal of bullet points and newlines and with numerals transcribed to words) is 60129~characters long, which equates to 11591~words and 547~turns.

\begin{table}
\centering
    \begin{tabular}{ll}
    \toprule
        Measure & Quantity \\
    \midrule
        Number of characters               & 60,129   \\
        Number of words                    & 11,591   \\
        Number of speaker turns            & 547     \\
        Audio book duration                & 113 min \\
        ASR dataset duration (speech only) & 79 min \\
    \bottomrule
    \end{tabular}
\end{table}

\vspace{0.7cm}

\section{Description of the data processing pipeline}
\label{sec:processing_description}

In this section we are describing the process of transforming the data obtained by the first author of the original book, and the last author of this paper, to obtain the final word-aligned computer-readable dataset, useful both for developing, adapting and evaluating speech technologies, as well as releasing the work in a digital form.


The processing involved the following steps:
\begin{description}
    \item[Chapter-based Segmentation] Audio and text were manually segmented into chapters, where chapter 00 denotes the preface and chapters 01 to 28 contain the body section of the book.
    \item[Voice Activity Detection] Every chapter was analyzed with a voice activity detection model~\cite{Bredin2020, Bredin2021}\footnote{\url{https://huggingface.co/pyannote/segmentation}} which automatically gives spans that contain human speech. Some chapters begin or end with music which would hinder downstream processing. By detecting the parts of the audio where only speech is present, the relevant data can be successfully processed in the downstream.
    \item[Trimming] For every chapter the audio was trimmed so that only the parts containing speech are preserved.
    \item[Diarisation] Intervals, where specific speakers are speaking, were identified with a diarisation model\footnote{\url{https://huggingface.co/pyannote/speaker-diarization-3.1}} \cite{Plaquet23,Bredin23}.
    \item[Exporting] For manual corrections and inspection, the trimmed audio and diarised data were exported in EXB format.
    \item[Manual interventions] EXB files were inspected in Exmaralda Partitur Editor\footnote{\url{https://exmaralda.org/en/partitur-editor-en/}}. Any misdiarised turns were manually labeled and automatically corrected afterwards. The speakers identified during diarisation, or added during the manual inspection, were labelled with their characters' names (e.g. Mići Princ, Pisac, Rožica,~\ldots).
    \item[Alignment] Texts were normalised (special dialect characters  \^{i}, \"{i}, \"{a}, \^{a}, and \"{e} substituted with analogs from standard Croatian, punctuation characters were removed, numerals were changed to words). The normalised texts were aligned with the audio using Kaldi~\cite{Povey_ASRU2011}, similar to the process recently used to word-align the Slovenian Gos corpus~\cite{verdonik2024gos}. In rare cases, additional manual interventions were performed on texts to assure successful alignment (e.g. \texttt{Exup\'{e}ry} was changed to \texttt{Eksuperi} for alignment and then reverted after successful processing, some transcript errors were also identified during that process and rectified). The resulting aligned data, each word from the transcript having the start and end timestamp in the recording, were encoded in a json and the Exmaralda EXB format.
    \item[Data transformation for ASR] With the entire Mići Princ diarised, aligned, and manually inspected, the construction of an ASR flavour of the dataset was possible. Since most modern ASR models require relatively short segments, the dataset was re-segmented so that the instances' duration is shorter than 30~seconds. Instances from chapters 13 and 15 were kept aside for constructing the testing subset. They feature two speakers that are very common in the book (Autor and Mići Princ), as well as two additional speakers, Geograf and Dilavac, who do not occur in the rest of the data at all, which allows for examining performance differences on new versus known speakers.
\end{description}

\section{Final encoding of the resulting dataset}
\label{sec:final_encoding}
The final encoding of the constructed dataset was uploaded to the CLARIN.SI FAIR repository\footnote{\url{http://hdl.handle.net/11356/1765}}
~\cite{clarin_mici_princ_repo}. The encoding consists of the following files:
\begin{description}
    \item[MP.wav.tgz] audio files in wav format, one file per chapter
    \item[MP.mp3.tgz] audio files in mp3 format, one file per instance in the ASR dataset
    \item[MP.json.tgz] verbatim dataset in JSON format, as described below in Subsection~\ref{subsec:mpjson}
    \item[MP.asr.json.tgz] ASR-specific dataset in JSON format, as described below in Subsection~\ref{subsec:mpasrjson}
    \item[MP.exb.tgz] dataset in EXB format, suitable for viewing in the Exmaralda Partitur Editor
    \item[speakers.json] a file with speaker metadata information, describing who translated and read parts for a specific character and the provenience of the speaker (name and wikidata link)
\end{description}

\subsection{MP.json encoding}
\label{subsec:mpjson}

In the JSON encoding of the verbatim dataset, containing all the available information, each JSON file covers one chapter. Each entry covers one speaker turn and contain the following attributes:
\begin{description}
    \item[speaker] Character who is speaking in the current turn
    \item[text] Original text, as it appears in the book, with no alterations except 1.)~numerals being written with words and 2.)~parts not pronounced in the audio book omitted.
    \item[char\_s] Character offset start, denoting how many characters from the start of the chapter in textual format the turn starts
    \item[char\_e] Character offset end, i.e., how many characters from the start of the chapter does the turn end in the text version of the chapter
    \item[time\_s] Temporal offset start, i.e., how many seconds after the start of the chapter recording the turn start
    \item[time\_e] Temporal offset end, i.e., how many seconds after the start of the chapter the turn ends in the recording
    \item[words] A list of key:value pairs for attributes \texttt{char\_s, char\_e, time\_s, time\_e} for individual words, i.e., information for each word where it is located in the textual version and the audio version of the dataset.
\end{description}

\subsection*{Example entry}

In the example entry below, we see a short turn of the \emph{Mići Princ} saying \emph{Prosin vas, narišite mi ovcu}. Futhermore, we know that in the textual version of the chapter we can find this turn between characters at indices 595 and 623. We also know that the spoken form of this turn can be found in the recording between seconds 102.87 and 104.92. Finally, for each of the words, we have similar offset information, for the first word, \emph{Prosin}, the text version being available between character indices 595 and 601, and its pronunciation between seconds 102.87 and 103.34.

\begin{center}
\fbox{\begin{minipage}{0.9\textwidth}
{\ttfamily\small
\obeylines
\{\\
\hspace*{2em}"char\_e": 623,\\
\hspace*{2em}"char\_s": 595,\\
\hspace*{2em}"speaker": "Mi\'ci Princ",\\
\hspace*{2em}"text": "Prosin vas, nari\v{s}ite mi ovcu!",\\
\hspace*{2em}"time\_e": 104.92,\\
\hspace*{2em}"time\_s": 102.87,\\
\hspace*{2em}"words": [\{\\
\hspace*{4em}"char\_e": 601,\\
\hspace*{4em}"char\_s": 595,\\
\hspace*{4em}"time\_e": 103.34,\\
\hspace*{4em}"time\_s": 102.87\\
\hspace*{2em}\},\\
\hspace*{2em}\{\\
\hspace*{4em}"char\_e": 605,\\
\hspace*{4em}"char\_s": 602,\\
\hspace*{4em}"time\_e": 103.59,\\
\hspace*{4em}"time\_s": 103.34\\
\hspace*{2em}\}, \dots]\\
\}}
\end{minipage}}
\end{center}

\subsection{MP.asr.json encoding}
\label{subsec:mpasrjson}

In this section the encoding of the ASR flavour of the dataset is described. It is much simpler than the verbatim encoding described in the previous chapter. Each json covers one chapter. Each entry covers speech in the length of up to 30 seconds. In case of chapters 13 and 15, the testing chapters, it is ensured that each turn is spoken by just one speaker. Each instance contains the following attributes:

\begin{description}
    \item[audio] Name of the audio file in MP.mp3.tgz
    \item[text] Text of the instance
    \item[normalized\_text] Text without bullet points and newlines, with special characters substituted
    \item[speaker] Character speaking the instance. This attribute is only present in the testing chapters 13 and 15.
\end{description}

The biggest changes in this ASR-flavoured version of the data are that 1.)~recording snippets are available in mp3 format for each instance, up to 30 seconds long, 2.)~there is no alignment information available, 3. text normalization was performed, with bullet points and newlines removed, and accented characters that do not appear in standard Croatian being substituted. With these three changes it is easy to produce instances of short speech and corresponding text snippets, as preferred by the ASR community.

To further boost the visibility of the dataset overall, and especially its application in ASR, this ASR version of the Mići Princ dataset was also published to HuggingFace dataset hub\footnote{\url{https://huggingface.co/datasets/classla/Mici_Princ}}, which enables adapting or evaluating speech technologies on this dataset in a few lines of code.

\subsection*{Example entry}

In the below example we can observe that each instance has an mp3 file attached (the file name encoding the chapter, as well as time offsets), and the text having a normalised version consisting only of standard Croatian characters.

\begin{center}
\fbox{\begin{minipage}{0.9\textwidth}
{\ttfamily\small
\obeylines
\{\\
\hspace*{2em}"audio": "MP\_13\_260.92-261.63.mp3",\\
\hspace*{2em}"text": "I to je s\"e?",\\
\hspace*{2em}"normalized\_text": "I to je se?"\\
\hspace*{2em}"speaker": "Mići Princ",\\
\}}
\end{minipage}}
\end{center}


\section{ASR experiments}
\label{sec:asr_experiments}

In this section we are describing our preliminary, but very successful ASR experiments on the dataset described in the previous sections. In the first subsection we are describing the setup of the ASR model fine-tuning procedure, the second subsection describes the overall evaluation of the resulting model, while a more detailed, speaker-specific analysis of the output is provided in the final subsection.

\subsection{Finetuning setup}

For the ASR technology we want to adapt to the Chakavian dataset we chose the \texttt{Whisper-large-v3}\footnote{\url{https://huggingface.co/openai/Whisper-large-v3}} \cite{radford2022whisper} model due to its reasonable\footnote{In \cite{Samardzic2024} Whisper-large-v3 is evaluated on a new Croatian ASR dataset, especially adapted to challenges in ASR (e.g. numbers being transcribed as numerals instead of words). In this setting Whisper-large-v3 outperformed other models and achieved character error rate as low as 6.68\% and word error rate of 16.18\%.} performance on standard Croatian language. In preliminary experiments, a brief hyper-parameter optimization was performed, in which 80~epochs and learning rate of 1e-5 were chosen as optimal, with effective batch size set to 16.

During fine-tuning with the chosen hyper-parameters, the model was serialized (i.e. saved to disk) every 4 epochs, so that various evaluations could be calculated post-festum as our fine-tuning was progressing.

\subsection{Evaluation}

The metrics used for evaluation are the two most standard metrics for ASR evaluation: word error rate (WER), which calculates the percentage of mistranscribed words, and character error rate (CER), the percentage of characters that were mistranscribed. Given that the metrics calculate the percentage of errors, lower values show better ASR performance. We use the implementation of the two metrics in the \texttt{evaluate}\footnote{\url{https://pypi.org/project/evaluate/}} package. Before calculating those metrics, both the model outputs and reference text is lower-cased and stripped of punctuation.

\begin{figure}
    \centering
    \includegraphics[width=\textwidth, alt={Two plots, one labelled WER and the other labelled CER. Both span from 0 epochs to 80 epochs horisontally, and from 0 to 0.5 on y-axis. The plots show WER and CER during finetuning with different colours for different speakers. Metrics for Mići Princ and Dilavac first drop, then rise dramatically, and then continue to drop. Other speakers mostly drop consistently during fine tuning.}]{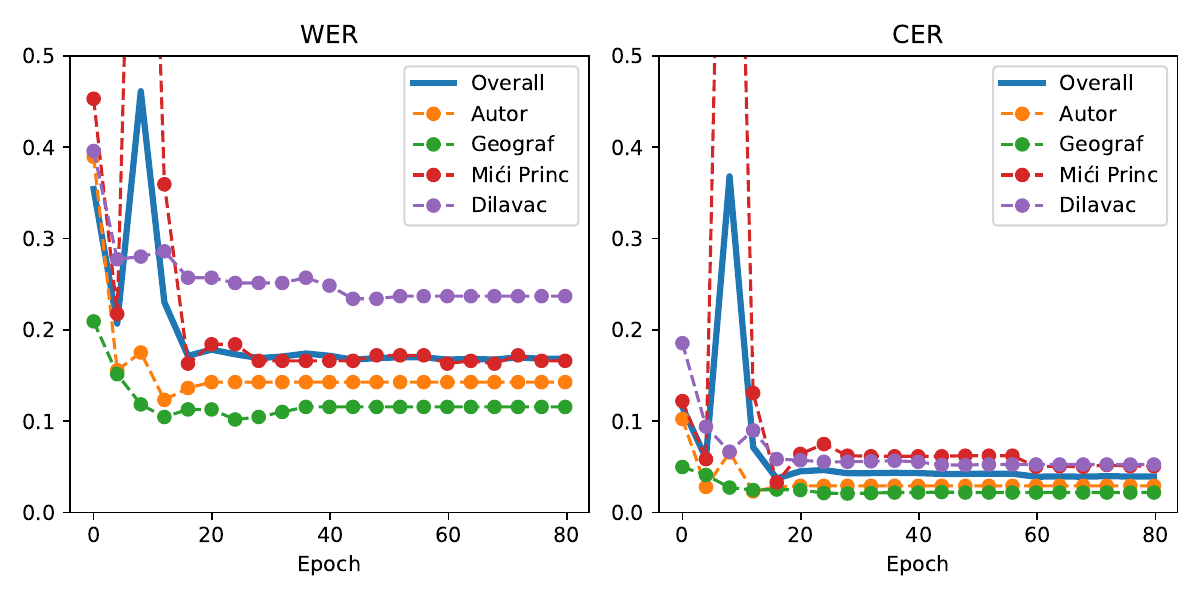}
    \caption{Metrics, achieved by the model during fine-tuning.}
    \label{fig:metrics}
\end{figure}

In Figure~\ref{fig:metrics} we present the development of both metrics, overall and by speaker, as the fine-tuning progresses through the 80 selected epochs (each instance in the dataset is used 80 times in fine-tuning). Both metrics exhibit the same profile during the fine-tuning process. The left-most datapoints, at epoch 0, on both plots show performance of `vanilla' Whisper, before any fine-tuning took place. It is evident that on both metrics and all speakers fine-tuning improved results visibly, which is especially important given that two out of four speakers, namely \emph{Dilavac} and \emph{Geograf}, were not seen by the model during fine-tuning.

The `Overall' curve on Figure~\ref{fig:metrics} shows a very much expected profile. In the first part of the training both metrics get worse, but after a while the model picks up on the specificities of the dialect and the metrics drop. An inspection of the outputs after just a few epochs shows for the drastic deterioriation of the output is due to hallucinations (many repetitions of predicted character sequences) the Whisper model is known for while being adapted with additional data.

Comparing the output of the model with the reference text on the test split showed the model to be well adapted to the dialects present in the dataset, with differences mostly due to:
\begin{itemize}
    \item incorrect segmentation (e.g. \texttt{zvizdami} being transcribed as \texttt{zvizda\textvisiblespace\phantom{}mi})
    \item regressing to standard language (e.g. \texttt{š njimi}$\rightarrow$\texttt{s njimi})
    \item minor differences, where audio could be transcribed in two ways, and in a dialect without orthography both versions could be considered valid (e.g. \texttt{diljivaš}$\leftrightarrow$\texttt{dilivaš})
\end{itemize}

In one case our model correctly identified a mismatch between the printed and audio versions of the book that we missed to rectify in our processing pipeline. Since the disagreement was indisputably evident from listening to the audio (\texttt{priti} versus \texttt{arivat}), the ASR dataset was corrected to better reflect the task at hand. This identified mistranscription, performed better by the model than by us, humans, shows: 1.)~that today's ASR has become very good and 2.)~that we might have additional minor issues in the data that we will have to improve for the second release of the data. What is important is that in the whole test set, this was the only mistranscription identified, which shows these mistranscriptions to be very infrequent, and thereby the dataset of high quality.

\subsection{Speaker-based error analysis}

In this subsection we look at the performance of the model as it is being fine-tuned on the basis of each speaker in our testing data, namely two speakers very well represented in the fine-tuning data, \emph{Autor} and \emph{Mići princ}, and two speakers not present in our fine-tuning data, namely \emph{Dilavac} and \emph{Geograf}.

At the beginning of the fine-tuning process, Autor, Mići Princ, and Dilavac perform worse than Geograf on both metrics. After the first serialization at 4 epochs into the training the metrics drop for all speakers, after which performance of some speakers keeps improving, while for other both error rates explode. This phenomenon was investigated by examining the outputs of the models, and wild hallucinations (mostly repetitions a single word at the end of the output) were found to be the root cause for the significantly increased error rates. For some insofar yet unexplained reason, speaker reading lines for Mići Princ seems to be the most affected by this phenomenon. After some additional fine-tuning, these hallucinations become much less frequent, yielding better and more accurate results.

One hypothesis about this speaker-dependent difference is that Mići Princ is the speaker most different to standard pronunciation on the word level (on epoch 0 it has the highest word error rate), and that its need for adaptation gets affected with hallucinations. This speaker is also the most frequent speaker in the test data, which makes the overall metric explode as well.

Not all speakers followed the same metrics profile during training. Speakers present in the training data, Mići Princ and Autor, suffer from the aforementioned performance drop in the beginning of the testing, while new speakers seem not to. Geograf quickly achieves optimal performance, whereas the metrics for Dilavac drop much later in the training. Another possible hypothesis for differences in behaviour is not (just) the initial performance, but also that speakers present in the fine-tuning data are especially prone to hallucinations (over-generation) until the model gets properly fine-tuned.

We have stated two hypotheses on the difference in per-user performance as fine-tuning progresses, testing any of these going outside the scope of this paper, and will therefore have to be inspected more thoroughly in future research.

Comparing the per-speaker metrics with the map from Figure~\ref{fig:map} shows no significant geographic trend. Dilavac and Geograf both live very far from the weighed average of training data, which lies just south of Mići Princ and Autor, yet they achieve the worst and the best metrics respectively on the majority of finetuned models. To properly address the search for the existance of geographic trends a much bigger dataset would be needed, where content-based differences would average out.

In Table~\ref{tab:metrics} we present the initial, epoch 0 evaluation (\emph{vanilla}) of the model on both metrics and per each speaker and overall. We compare this Whisper-v3-large-before-adaptation performance with the final performance of the model at epoch 80 (\emph{fine-tuned}). We also report the relative error reduction, which encodes the percentage of errors that were successfully removed from the output of the system with our model adaptation through model fine-tuning.

After 80 epochs the model reaches CER of 3.95\% and WER 16.83\%, which are very good numbers for the complexity of the underlying problem. What is most important, similar numbers can be observed on the previously unseen speakers, which shows the generality of our adaptation. However, one still has to bear in mind that these are studio-recorded spoken utterances and that transcribing speech in less controlled environments would quite likely be much more challenging.

As expected, the relative error reduction of word error rate on the speakers seen during fine-tuning is higher (63.32\% and 63.33\%) than for the two unseen speakers (44.75\% and 40.15\%). This trend, however, does not hold for character error rate, where the overall largest improvement is measure with the \emph{Dilavac} character, which is speaking in a heavy dialect, with a very high character error rate of the vanilla model of 18.55\%, shrinking with the adaptation to 5.27\%, thereby 71.59\% of the error on character level being removed.

We can overall report very good results due to adaptation, with 52.5\% of error being removed on the level of words, and 65.77\% being removed on the level of characters.

The final model was published on HuggingFace model hub\footnote{\url{https://huggingface.co/classla/Whisper-large-v3-mici-princ}}, hoping to increase visibility of the dataset it has been fine-tuned on, but also to motivate future data-driven projects on this and other dialects.

\begin{table}
    \centering
    \begin{subtable}{\textwidth}
        \centering
        \begin{tabular}{lccc}
            \toprule
            \multicolumn{1}{c}{\textbf{speaker}} & \multicolumn{1}{c}{\textbf{vanilla}} & \multicolumn{1}{c}{\textbf{finetuned}} & \multicolumn{1}{c}{\textbf{relative error reduction}} \\
            \midrule
            all                                  & 35.43\%                              & 16.83\%                                & 52.50\%                                               \\
            Autor                                & 38.96\%                              & 14.29\%                                & 63.32\%                                               \\
            Geograf                              & 20.94\%                              & 11.57\%                                & 44.75\%                                               \\
            Mići Princ                           & 45.32\%                              & 16.62\%                                & 63.33\%                                               \\
            Dilavac                              & 39.60\%                              & 23.70\%                                & 40.15\%                                               \\
            \bottomrule
        \end{tabular}
        \caption{Word error rate (WER)}
        \label{tab:WER}
    \end{subtable}
    \begin{subtable}{\textwidth}
        \centering
        \begin{tabular}{lccc}
            \toprule
            \multicolumn{1}{c}{\textbf{speaker}} & \multicolumn{1}{c}{\textbf{vanilla}} & \multicolumn{1}{c}{\textbf{finetuned}} & \multicolumn{1}{c}{\textbf{relative error reduction}} \\
            \midrule
            all                                  & 11.54\%                              & 3.95\%                                 & 65.77\%                                               \\
            Autor                                & 10.24\%                              & 2.93\%                                 & 71.39\%                                               \\
            Geograf                              & 4.99\%                               & 2.19\%                                 & 56.11\%                                               \\
            Mići Princ                           & 12.21\%                              & 5.09\%                                 & 58.31\%                                               \\
            Dilavac                              & 18.55\%                              & 5.27\%                                 & 71.59\%                                               \\
            \bottomrule
        \end{tabular}
        \caption{Character error rate (CER)}
        \label{tab:CER}
    \end{subtable}
    \caption{Breakdown of metrics achieved with vanilla (\texttt{Whisper-large-v3}) and the finetuned model.}
    \label{tab:metrics}
\end{table}
\clearpage

\section{Limitations}
\label{sec:limitations}

There is a series of limitations that we want to put forward.

The frequency of special characters (e.g. \"{e}, \"{i}, \"{a}) is low, the most common occurs 29 times out of the total 60,129 characters in the dataset, the least common only appears once. With this in mind we omitted modelling them with our ASR model, which is a limitation of this approach, but with so few occurences of so many special characters we feel the introduction of them would only render the model less reliable.

As with the aforementioned \texttt{priti}$\leftrightarrow$\texttt{arivat} example, it is possible that there are other discrepancies between the audio and the printed version of the book. However, we expect for such potential discrepancies to be very infrequent, given that we were able to find just one in all of the test data.

Our hyper-parameter search was by no means exhaustive, and it is possible that a better fine-tuning setup could exist. Additionally, in our hyper-parameter search we used the same data for training and evaluating as we did for fine-tuning itself, which is not the best practice.

Finally, while our error reductions, as well as overall measured performance is very reassuring, we must stress that this evaluation was performed on acted speech, recorded in a studio setting. Any dialectal speech production out in the wild will surely be much more challenging.

\section{Conclusions}
\label{sec:conclusions}

In this paper we have presented our efforts in ensuring the usefulness of two traditional releases, a printed book, and an audio book, both being a translation of \emph{The Little Prince} into Chakavian micro-dialects, beyond these two traditional means of publication.

The first use case for the new dataset, one we have already followed in this paper, is the adaptation of an automatic speech recognition system to the Chakavian dialect. Similar usage can be expected in the future as well, with the dataset becoming both a fine-tuning and an evaluation dataset for future models.

Another use case is the application of data in dialectal research, although the data are acted, so caution is needed for such data usage. However, given the absolute lack of open dialectal data for research, we consider this dataset to improve the data landscape on this front as well.

The third use case that we very much hope for is the preparation of a digital online edition of the translation, where audio and text content could be followed in parallel. Our own experience with the content is that neither the textual nor the audio content is informative enough to delve deep in the rich and aesthetically pleasing content available in the two separate traditional releases.

With the first use case we have illustrated the feasibility of adapting existing tools and frameworks for standard languages to either dialects or other related languages with little resources. In the process two datasets were compiled and published, one following closely the structure of the Mići Princ printed book and audiobook, and the second, compiled with specific ASR applications in mind.

During the ASR system fine-tuning process we noticed interesting disadvantageous transient phenomena, mostly overgeneration of the final text, but after a long enough fine-tuning process, the output is stable with little bias towards new speakers.

We hope that this project will be motivation for further similar endeavours where content right holders will be open for technology-savvy language and speech preservation enthusiasts to encode their work under an open license for the benefit of all involved parties, as well as society as a whole.

\section*{Acknowledgements}

    This work was partially funded by the programme P6-0411 ``Language Resources and Technologies for Slovene'',  the CLARIN.SI infrastructure, and the project J7-4642 ``MEZZANINE -- Development of Spoken Language Resources and Speech Technologies for the Slovenian Language'', all financed by the Slovenian Research and Innovation Agency (ARIS).

    We would very much like to thank the following organisations and individuals: the PEEK\&POKE museum for allowing, together with the final author, Tea Perinčić, for these two traditional releases to be made available in a new, AI-friendly format, under a permissive license; John Scott, Marko Simonović and Keith Langston for making the first author aware of texual and audio releases in the Chakavian dialect.

\bibliographystyle{apacite}
\bibliography{references}

@inproceedings{Povey_ASRU2011,
  author    = {Povey, Daniel and Ghoshal, Arnab and Boulianne, Gilles and Burget, Lukas and Glembek, Ondrej and Goel, Nagendra and Hannemann, Mirko and Motlicek, Petr and Qian, Yanmin and Schwarz, Petr and Silovsky, Jan and Stemmer, Georg and Vesely, Karel},
  month     = dec,
  title     = {The Kaldi Speech Recognition Toolkit},
  booktitle = {IEEE 2011 Workshop on Automatic Speech Recognition and Understanding},
  year      = {2011},
  publisher = {IEEE Signal Processing Society},
  location  = {Hilton Waikoloa Village, Big Island, Hawaii, US},
  note      = {IEEE Catalog No.: CFP11SRW-USB}
}

@inproceedings{Plaquet23,
  author    = {Alexis Plaquet and Hervé Bredin},
  title     = {{Powerset multi-class cross entropy loss for neural speaker diarization}},
  year      = 2023,
  booktitle = {Proc. INTERSPEECH 2023}
}

@inproceedings{Bredin23,
  author    = {Hervé Bredin},
  title     = {{pyannote.audio 2.1 speaker diarization pipeline: principle, benchmark, and recipe}},
  year      = 2023,
  booktitle = {Proc. INTERSPEECH 2023}
}

@inproceedings{Bredin2021,
  title     = {{End-to-end speaker segmentation for overlap-aware resegmentation}},
  author    = {{Bredin}, Herv{\'e} and {Laurent}, Antoine},
  booktitle = {Proc. Interspeech 2021},
  address   = {Brno, Czech Republic},
  month     = {August},
  year      = {2021}
}

@inproceedings{Bredin2020,
  title     = {{pyannote.audio: neural building blocks for speaker diarization}},
  author    = {{Bredin}, Herv{\'e} and {Yin}, Ruiqing and {Coria}, Juan Manuel and {Gelly}, Gregory and {Korshunov}, Pavel and {Lavechin}, Marvin and {Fustes}, Diego and {Titeux}, Hadrien and {Bouaziz}, Wassim and {Gill}, Marie-Philippe},
  booktitle = {ICASSP 2020, IEEE International Conference on Acoustics, Speech, and Signal Processing},
  address   = {Barcelona, Spain},
  month     = {May},
  year      = {2020}
}

@misc{radford2022whisper,
  doi       = {10.48550/ARXIV.2212.04356},
  url       = {https://arxiv.org/abs/2212.04356},
  author    = {Radford, Alec and Kim, Jong Wook and Xu, Tao and Brockman, Greg and McLeavey, Christine and Sutskever, Ilya},
  title     = {Robust Speech Recognition via Large-Scale Weak Supervision},
  publisher = {arXiv},
  year      = {2022},
  copyright = {arXiv.org perpetual, non-exclusive license}
}

@misc{clarin_mici_princ_repo,
  title     = {The "{M}i{\'c}i {P}rinc" text and speech dataset of {C}hakavian micro-dialects},
  author    = {Ljube{\v s}i{\'c}, Nikola and Rupnik, Peter and Perin{\v c}i{\'c}, Tea},
  url       = {http://hdl.handle.net/11356/1765},
  note      = {Slovenian language resource repository {CLARIN}.{SI}},
  copyright = {Creative Commons - Attribution-{ShareAlike} 4.0 International ({CC} {BY}-{SA} 4.0)},
  issn      = {2820-4042},
  year      = {2024}
}

@book{mici_princ_the_book,
  title     = {Mići Princ},
  author    = {Saint-Exupéry, Antoine de},
  editor    = {Perinčić, Tea},
  address   = {Rijeka},
  publisher = {Udruga Calculus, Muzej informatike “Peek\&Poke”},
  year      = {2021},
  note      = {Prijevod djela: Le petit prince.}
}

@inproceedings{Samardzic2024,
  title     = {Mak na konac: A multi-reference Speech-to-text Benchmark for Croatian and Serbian},
  author    = {Tanja Samard{\v z}i{\'c} and Peter Rupnik and Mirjana Starovi{\'c} and Nikola Ljubeši{\'c}},
  booktitle = {Proceedings of the {L}anguage {T}echnologies and {D}igital {H}umanities 2024 conference ({JT-DH} 2024)},
  year      = {2024},
  address   = {Ljubljana, Slovenia},
  month     = sep,
  day       = {19--20}
}

@inproceedings{kantharuban-etal-2023-quantifying,
  title     = {Quantifying the Dialect Gap and its Correlates Across Languages},
  author    = {Kantharuban, Anjali  and
               Vuli{\'c}, Ivan  and
               Korhonen, Anna},
  editor    = {Bouamor, Houda  and
               Pino, Juan  and
               Bali, Kalika},
  booktitle = {Findings of the Association for Computational Linguistics: EMNLP 2023},
  month     = dec,
  year      = {2023},
  address   = {Singapore},
  publisher = {Association for Computational Linguistics},
  url       = {https://aclanthology.org/2023.findings-emnlp.481},
  doi       = {10.18653/v1/2023.findings-emnlp.481},
  pages     = {7226--7245}
}

@misc{zhao2023survey,
  title         = {A Survey of Large Language Models},
  author        = {Wayne Xin Zhao and Kun Zhou and Junyi Li and Tianyi Tang and Xiaolei Wang and Yupeng Hou and Yingqian Min and Beichen Zhang and Junjie Zhang and Zican Dong and Yifan Du and Chen Yang and Yushuo Chen and Zhipeng Chen and Jinhao Jiang and Ruiyang Ren and Yifan Li and Xinyu Tang and Zikang Liu and Peiyu Liu and Jian-Yun Nie and Ji-Rong Wen},
  year          = {2023},
  eprint        = {2303.18223},
  archiveprefix = {arXiv},
  primaryclass  = {cs.CL}
}

@inproceedings{ljubesic-etal-2022-parlaspeech,
  title     = {{P}arla{S}peech-{HR} - a Freely Available {ASR} Dataset for {C}roatian Bootstrapped from the {P}arla{M}int Corpus},
  author    = {Ljube{\v{s}}i{\'c}, Nikola  and
               Kor{\v{z}}inek, Danijel  and
               Rupnik, Peter  and
               Jazbec, Ivo-Pavao},
  editor    = {Fi{\v{s}}er, Darja  and
               Eskevich, Maria  and
               Lenardi{\v{c}}, Jakob  and
               de Jong, Franciska},
  booktitle = {Proceedings of the Workshop ParlaCLARIN III within the 13th Language Resources and Evaluation Conference},
  month     = jun,
  year      = {2022},
  address   = {Marseille, France},
  publisher = {European Language Resources Association},
  url       = {https://aclanthology.org/2022.parlaclarin-1.16},
  pages     = {111--116}
}

@misc{11356/1914,
  title     = {Parliamentary spoken corpus of Croatian {ParlaSpeech}-{HR} 2.0},
  author    = {Ljube{\v s}i{\'c}, Nikola and Kor{\v z}inek, Danijel and Rupnik, Peter},
  url       = {http://hdl.handle.net/11356/1914},
  note      = {Slovenian language resource repository {CLARIN}.{SI}},
  copyright = {Creative Commons - Attribution-{ShareAlike} 4.0 International ({CC} {BY}-{SA} 4.0)},
  issn      = {2820-4042},
  year      = {2024}
}

@article{kuvavc2016croatian,
  title     = {Croatian adult spoken language corpus (HrAL)},
  author    = {Kuva{\v{c}} Kraljevi{\'c}, Jelena and Hr{\v{z}}ica, Gordana},
  journal   = {FLUMINENSIA: {\v{c}}asopis za filolo{\v{s}}ka istra{\v{z}}ivanja},
  volume    = {28},
  number    = {2},
  pages     = {87--102},
  year      = {2016},
  publisher = {Odsjek za kroatistiku Filozofskoga fakulteta Sveu{\v{c}}ili{\v{s}}ta u Rijeci}
}

@misc{kovavcevic2002croatian,
  title  = {Croatian corpus, CHILDES},
  author = {Kova{\v{c}}evi{\'c}, Melita},
  year   = {2002}
}

@inproceedings{verdonik2024gos,
  title     = {Gos 2: A New Reference Corpus of Spoken Slovenian},
  author    = {Verdonik, Darinka and Dobrovoljc, Kaja and Erjavec, Toma{\v{z}} and Ljube{\v{s}}i{\'c}, Nikola},
  booktitle = {Proceedings of the 2024 Joint International Conference on Computational Linguistics, Language Resources and Evaluation (LREC-COLING 2024)},
  pages     = {7825--7830},
  year      = {2024}
}

@inproceedings{ljubesic-etal-2024-dialect,
  title     = {{DIALECT-COPA}: Extending the Standard Translations of the {COPA} Causal Commonsense Reasoning Dataset to South Slavic Dialects},
  author    = {Ljube\v{s}i\'{c}, Nikola and Galant, Nada and Ben\v{c}ina, Sonja and \v{C}ibej, Jaka and Milosavljevi\'{c}, Stefan and Rupnik, Peter and Kuzman, Taja},
  editor    = {Scherrer, Yves  and
               Jauhiainen, Tommi  and
               Ljube{\v{s}}i{\'c}, Nikola  and
               Nakov, Preslav  and
               Tiedemann, J{\"o}rg  and
               Zampieri, Marcos},
  booktitle = {{E}leventh {W}orkshop on {NLP} for {S}imilar {L}anguages, {V}arieties and {D}ialects ({VarDial} 2024)},
  year      = {2024},
  address   = {Mexico City, Mexico},
  publisher = {Association for Computational Linguistics}
}
\end{document}